\documentclass[conference]{IEEEtran}
\IEEEoverridecommandlockouts
\usepackage{cite}
\usepackage{amsmath,amssymb,amsfonts}
\usepackage{algorithmic}
\usepackage{graphicx}
\usepackage{textcomp}
\usepackage{xcolor}

\usepackage{setspace}
\usepackage{multirow}
\usepackage{enumitem}

\begin{document}

\title{Awareness of Secure Coding Guidelines in the Industry - A first data analysis}

\author{\IEEEauthorblockN{Tiago Espinha Gasiba}
\IEEEauthorblockA{\textit{Siemens AG} \\
Munich, Germany \\
tiago.gasiba@siemens.com}
\and
\IEEEauthorblockN{Ulrike Lechner}
\IEEEauthorblockA{\textit{Universität der Bundeswehr}\\
\textit{München} \\
Munich, Germany \\
ulrike.lechner@unibw.de}
\and
\IEEEauthorblockN{Maria Pinto-Albuquerque}
\IEEEauthorblockA{\textit{Instituto Universitário de} \\
\textit{Lisboa (ISCTE-IUL), ISTAR-IUL} \\
Lisboa, Portugal \\
maria.albuquerque@iscte-iul.pt}
\and
\IEEEauthorblockN{Daniel Mendez Fernandez}
\IEEEauthorblockA{\textit{Blekinge Institute of Technology} \\
Karlskrona, Sweden \\
daniel.mendez@bth.se}
}

\maketitle

\begin{abstract}
Software needs to be secure, in particular, when deployed to critical infrastructures. Secure coding guidelines capture practices in industrial software engineering to ensure the security of code. This study aims to assess the level of awareness of secure coding in industrial software engineering, the skills of software developers to spot weaknesses in software code, avoid them, and the organizational support to adhere to coding guidelines. The approach draws on well-established theories of policy compliance, neutralization theory, and security-related stress and the authors' many years of experience in industrial software engineering and on lessons identified from training secure coding in the industry. The paper presents the questionnaire design for the online survey and the first analysis of data from the pilot study.
\end{abstract}

\begin{IEEEkeywords}
Security, Secure Coding, Software Development, Best Practices, Security Awareness, Industry
\end{IEEEkeywords}

\section{Introduction}
\label{sec:introduction}
Software for critical infrastructures needs to be secure. The cyber-attacks to various industry sectors~\cite{WWW_CVE} are ever-increasing in their number, variety, and consequences. These consequences range from loss-of-life, loss-of-business (e.g. through service disruption), loss-of-confidential data (through sensitive information leakage), data-compromise (though unauthorized changes) and monetary losses~\cite{EternalBlue_FinancialImpact}. This problem is made clear through the number of alerts and advisories issued yearly by the Industrial Control System - Computer Emergency Team (ICS-CERT) from the United States Department of Homeland Security~\cite{ICS_CERT}.
Before 2014 less than 100 advisories per year have been issued, while from 2017 to 2019 more than 200 advisories per year have been issued.
This increase underpins the need for secure coding practices for software deployed to industrial (critical) infrastructures. Another argument for more efforts to increase the security of code is the Common Vulnerabilities and Exposures (CVE) \cite{WWW_CVE} online database, with the relationship between existing vulnerabilities and Common Weaknesses Enumeration (CWE) \cite{WWW_CWE_2019}.
This online database shows that the number of known vulnerabilities has more than doubled from 2016 to 2017 (from 6447 to 14714). Also, on average, about 2000 new vulnerabilities per year have been added since then.
In a recent (2019) GitLab survey with more than 4000 software developers, Patel et al.~\cite{gitlab_2019} found that less than 50\% of software developers can identify secure coding vulnerabilities. Secure coding based on secure coding guidelines (SCG) is one way to create secure code in industrial software engineering.

The study at hand is part of a design research effort: designing a serious game to raise awareness for secure coding and increase knowledge of secure coding guidelines among software developers~\cite{gasiba_re19}. The plan is to collect data on awareness of vulnerabilities and training levels of industrial software developers and, in particular, on why do software developers comply with secure coding guidelines. In this study, we are interested in understanding how well secure software development practices are established in the industry, the factors that influence whether industrial software engineers comply with secure coding guidelines, and the knowledge of software developers of secure coding guidelines. 

The survey design is based on theories about security policy compliance \cite{moody2018toward,bulgurcu2010information}, security-related stress \cite{d2014understanding}, neutralization theory \cite{siponen2010neutralization} and information on typical vulnerabilities. Furthermore, it is based on the first authors' experience in teaching secure coding in different programming languages and vulnerabilities in industrial software engineering.

The results of the survey will impact both theory and practice. By understanding the reasons that lead to a lack of awareness by software developers, it is possible to better tailor a serious game \cite{2016_Doerner_Serious_Games} and motive the overall research approach of designing a serious game for secure coding training for industrial software engineers. This paper presents and motivates our survey design and first data analysis of data collected in the extensive process to develop the questionnaire. We focus our results on the core questions that remained stable throughout the survey's development, and that made to the final survey.

This paper is organized as follows.
In section~\ref{sec:related_work}, we present related work.
Section~\ref{sec:design} gives details the research design.
This section also describes how the survey was piloted in three different companies.
Section~\ref{sec:results} shows the results of our research.
In particular, it gives details on the selected theories for the survey, contributed questions, and offers an outline of the survey design.
Furthermore, it presents preliminary results from the survey piloting.
Lastly, it presents a comparison with previous work, threats to validity, and shortly discusses the work's impact.
Section~\ref{sec:conclusions} concludes this paper with a summary of the present study and outlines further work.
\section{Related work}
\label{sec:related_work}
Several previous studies have indicated that software developers lack secure programming skills.
In 2020, Bruce Schneier, a well-known security researcher and evangelist has stated that less than 50\% of software developers can spot security vulnerabilities in software~\cite{Schneier2020} - this is, unfortunately, not a new trend.
In 2011, Xie et al.~\cite{Xie2011} did several interviews with 15 senior professional software developers in the industry with an average of 12 years of experience. Their study has shown a disconnect between software security concepts and the role that the participants have in their jobs.

There are various studies on the search processes and quality of information on secure coding topics online, e.g., on platforms such as stack-overflow.
In the large scale study, Yang et al.~\cite{Yang2016} identify questions related to security that software developers ask on the platform stack-overflow.com. They conclude that software developers' questions could be categorized in several different topics, whereby they found more than 600 different items for every problem. Fisher et al~\cite{fischer2017stack} have shown that typical online platforms that software developers use to clarify development questions can be considered harmful, as the answers present in such platforms are not curated for correctness of security. Their work indicates that severe problems can arise if software developers use these references and are not aware of secure coding practices.
Acar et al.~\cite{Acar2017} did an extensive analysis of existing online resources that software developers can access to get information on secure programming issues. They found out that the quality of information is not guaranteed due to, e.g., outdated information, no concrete examples or exercises. The analyses illustrate that software developers need secure coding skills, as software developers cannot depend on the well-known online sources for secure coding topics.

Gasiba et al.~\cite{Gasiba2019_Raising,gasiba_re19} propose a method based on Capture-the-Flag events to train secure programming for software developers in the industry.
Votipka et al.~\cite{votipka2018toward} also discuss Capture-the-Flag events as a means to improve secure software development.
In~\cite{Davis2014}, Davis et al. discuss the benefits of Capture-the-Flag (CTF) for software developers and Graziotin et al.~\cite{2018_Graziotin_Happy_Developers} argue that \textit{happy developers are better coders}, i.e., write higher quality code \cite{ISO250xx}.

There are very few empirical results on the extent to which software developers comply with secure coding guidelines and practices. A notable recent study by Assal et al. \cite{Assal2019} analyses how developers influence and are influenced by secure coding processes.
They conclude that software developers are \textit{not the weakest link} and are very motivated towards software security. However, they do not cover deeply the reasons that lead software developers to comply or not comply to secure coding guidelines.

In the present work, the concept of awareness or IT-security awareness is used to conceptualize the knowledge or skills in the IT-security domain. Benenson et al. provide a literature review on IT-security awareness \cite{2014_Benenson_Defining_Security_Awareness}. Other conceptualizations that address compliance with IT-security policies or guidelines are given by Bulgurcu et al. \cite{bulgurcu2010information}. The Unified Model by Moody et al. \cite{moody2018toward} synthesizes research on security policy compliance.
Siponen et al. \cite{siponen2010neutralization} address possible reasons why software developers might discard the usage of secure coding guidelines.
Finally, D'Arcy et al. \cite{d2014understanding} use coping theory to explore the relationship between stress and deliberate policy violations.

\section{Research Design}
\label{sec:design}
To assess the usage of secure coding guidelines in an industrial setting, we designed a survey aimed at software developers in the industry. The questionnaire is designed under the banner of three research questions:
\begin{itemize}[leftmargin=+.385in]
    \item[ {\bf RQ1:}] How well established are secure software development practices in the industry?
    \item[ {\bf RQ2:}] Which factors lead software developers to use or ignore secure coding guidelines?
    \item[ {\bf RQ3:}] How well are software developers aware of secure coding guidelines?
\end{itemize}

\begin{figure*}[ht]
    \centering
    \begin{minipage}{.95\textwidth}
        \centering
        \includegraphics[width=.9\textwidth]{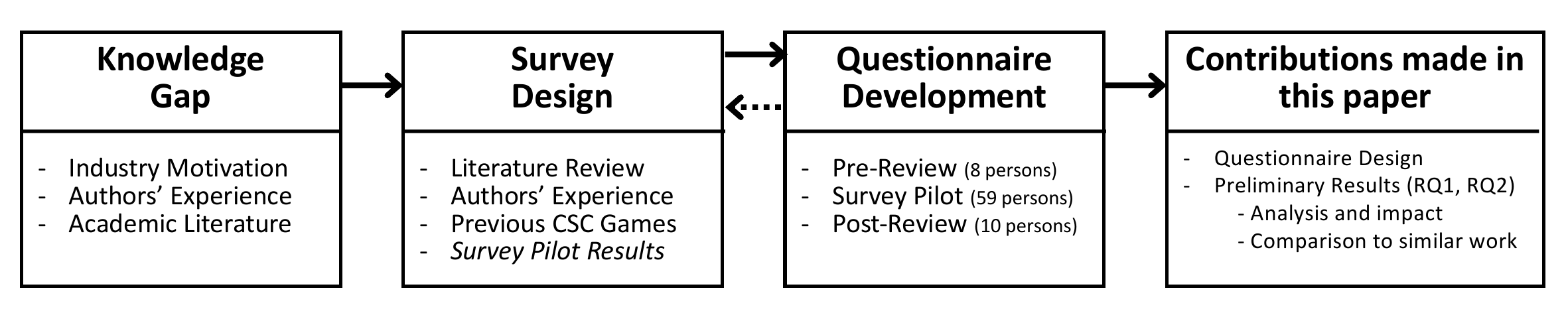}
        \vspace{-2.2em}
        {\\\hspace{\textwidth}
              {\scriptsize {\bf CSC}: CyberSecurity Challenges, {\bf RQ}: Research Question ~~~~~~~~~~~~~~~~~~~~~~~~~~~~~~~~~~~~~~~~~~~~~~~~~~~~~~~~~~~~~~~~~~~~~~~~~~~~~~~~~~~~~~~~~~~~~~~~~~~~~~~~~~~ } }
        \caption{Research Process}
        \label{fig:survey:development_method}
    \end{minipage}
\end{figure*}

These questions are driven by the industry's emerging need to identify factors that lead software developers to ignore secure coding guidelines in daily work and to which no answers can currently be found in previous scientific work.
Fig.~\ref{fig:survey:development_method} shows a summary of the systematic steps that we took in order to achieve this.
These steps are the following: 1) definition of the research goals through the knowledge gap, 2) survey design, 3) questionnaire development, and 4) contributions made in this paper.
In the following, we give details on each of these steps.

Our design of the questionnaire on secure coding guidelines draws from the first author's experience in industrial software engineering, from published experience reports in survey design~\cite{wagner2020}, and data collected in the context of a serious game~\cite{2016_Doerner_Serious_Games}. This information has been used for crafting the questions on knowledge on weaknesses in relevant programming languages.

We use the concept of IT security awareness with the three dimensions of knowledge following \cite{2014_Benenson_Defining_Security_Awareness}: (1) recognize the threat, (2) know solutions to the threat, and (3) act right. For the factors that influence compliance with secure coding guidelines, we use from the literature on IT security compliance \cite{moody2018toward,bulgurcu2010information}, neutralization theory \cite{siponen2010neutralization} and security-related stress \cite{d2014understanding}.
In the development of the questionnaire, we took an iterative approach with first a pre-review to eliminate simple errors and review the questionnaire structure. In this phase, we have collected review comments from 8 security experts that covered, in particular, the relevance of the questions (in particular the ones on programming languages and typical vulnerabilities as well as on the knowledge of secure coding standards). In the subsequent pilot, we collected responses from 59 software developers from the industry. The survey pilot was administered to participants in the context of a CyberSecurity Challenges (CSC) workshop on secure coding guidelines. Each workshop had a duration of three days. Table~\ref{tab:survey:pilot} shows a summary of the different companies where these workshops took place and where the survey pilot results were collected.
This work presents an analysis of the data collected from this phase, i.e. our results are based on the survey design's first iteration, as shown in figure \ref{fig:survey:development_method}.
After this phase, an additional post-review with ten security experts took place. The goal of this additional iteration was to refine and prepare a final version of the survey for a large-scale deployment based on the preliminary results hereby presented.
In particular, the preliminary results analysis led to the additional formulation of RQ3 at this stage.
However, in this paper, we do not present results concerning RQ3, since the data to tackle it is not available at the time of publication, but we present the survey's final design as an outcome of our work.
After completing a large-scale deployment of the final survey, the authors intend to present an in-depth analysis of these results in a forthcoming publication.
The process of questionnaire development was done Q4 in 2019 to April 2020. An essential inspiration for the refinement of the questionnaire was a recently published study on the interplay between developers and software security processes by Assal et al.~\cite{Assal2019}. Dillman et al.~\cite{dillman2011mail} was used as a guide to the questionnaire development process. The answers to the questionnaires' questions use the well-known Likert scale~\cite{likert1932technique}.

\begin{spacing}{.9}
  \begin{table*}[htb]
    \renewcommand{\arraystretch}{1.82}
    \footnotesize
    \centering
    \caption{Survey piloting according to delivered secure coding workshops}
    \begin{tabular}{|p{0.7cm}|p{1.7cm}|p{1.3cm}|p{4cm}|p{1.8cm}|}
      \hline
      \textbf{Pilot}   &
      \textbf{Company}   &
      \textbf{Region}    &
      \textbf{~~~~~~~Nr. of Developers}    &
      \textbf{~~~~When}
      \\ \hline
      \hline
      ~~\#1 &
      Company A  &
      ~~~China &
      30: 15 C/C\texttt{++}, 15 Web &
      ~~Aug. 2019
      \\ \hline
      ~~\#2 &
      Company B  &
      Germany &
      ~~7:  7 Web &
      ~~Sep. 2019
      \\ \hline
      ~~\#3 &
      Company C  &
      ~~Turkey &
      22: 22 C/C\texttt{++} &
      ~~Oct. 2019
      \\ \hline
    \end{tabular}
    \label{tab:survey:pilot}
  \end{table*}
\end{spacing}

\vspace{.5em}
The pilot study participants were comprised of 50\% software developers with more than five years of work experience and 50\% with an average of 3 years of work experience. The software developers work for companies that develop software for critical infrastructures. The participants came from China (51\%), Turkey (37\%), and Germany (12\%), and the ages ranged from 25 to 60.
On average, the participants have attended $1.6$ training events related to security over the last five years, whereby 6 participants have attended more than five secure coding related training events over the past five years.
The majority of the developers (68\%) are embedded developers, working in C/C\texttt{++} (62\%), and 38\% work regularly with other programming languages such as e.g., Java, Python. Data analysis of the pilot study was done by standard statistical methods using RStudio $1.2.5019$.

\section{Results}
\label{sec:results}
In this section, we present the selected base theories and also the final design of our survey.
We also provide a preliminary analysis of the results obtained during the pilot phase.

\subsection{The Questionnaire}
\begin{spacing}{.9}
  \begin{table*}[htb]
    \renewcommand{\arraystretch}{1.52}
    \footnotesize
    \centering
    \caption{Selected theories}
    \begin{tabular}{|p{0.7cm}|p{3.8cm}|p{1.4cm}|p{8.5cm}|}
      \hline
      \textbf{Type}    &
      \textbf{Name}    &
      \textbf{Ref.}    &
      \textbf{Description}
      \\ \hline
      \hline
      PC &
      Policy Compliance Theory  &
      \cite{bulgurcu2010information, moody2018toward} &
      Assess to which extent do the participants intend to comply with secure coding guidelines in their organization
      \\ \hline
      NT &
      Neutralization Theory  &
      \cite{siponen2010neutralization} &
      Determine which reasons do software developers mostly use in order not to comply with secure coding guidelines
      \\ \hline
      SRS &
      Security-Related Stress  &
      \cite{d2014understanding} &
      Determine to which extend does complying to secure coding guidelines lead to an increase of stress at work
      \\ \hline
      AW &
      Dimensions of Awareness  &
      \cite{2014_Benenson_Defining_Security_Awareness} &
      Measure awareness of secure coding guidelines by the software developers on perception, behavior and protection
      \\ \hline
    \end{tabular}
    \centering
   {\\\hspace{\textwidth}\scriptsize {\bf Ref}: Literature reference~~~~~~~~~~~~~~~~~~~~~~~~~~~~~~~~~~~~~~~~~~~~~~~~~~~~~~~~~~~~~~~~~~~~~~~~~~~~~~~~~~~~~~~~~~~~~~~~~~~~~~~~~~~~~~~~~~~~~~~~~~~~~~~~~~~~~~~~~~~~~~~~~~~~~~~~~~~~}
    \label{tab:survey:theories}
  \end{table*}
\end{spacing}
Table~\ref{tab:survey:theories} summarizes theories identified in the design of the questionnaire, through literature research, and also provides a brief description of the theory which motivates the inclusion in the questionnaire.  
Table~\ref{tab:survey:questions} gives details on the survey questions with particular emphasis on the mapping towards our research questions.
In this table, the type of question indicates the used established theory or a type of contributed question.
The reference and construct give details on the origin of the question. Column "survey question" presents the final adapted question.

The questionnaire comprises the following four sections: 1) demographic data, 2) secure coding awareness, 3) secure coding compliance, and 4) deterrents to compliance. 
The first part of the questionnaire includes general demographic questions on work experience, previous training on secure coding, the primary programming language used at work, used secure coding processes in the company, and software developed method.

The second section of the questionnaire deals with awareness for secure coding. This part is individualized according to the primary programming language selected in the first section.
For each example, three questions of high-impact vulnerabilities (according to~\cite{WWW_CWE_2019}) related to secure coding guidelines are presented, corresponding to Per1, Be1, and Prot1 in Table~\ref{tab:survey:questions}.
The theoretical background of these questions is the concept of IT security awareness, as defined by Hänsch et al.~\cite{2014_Benenson_Defining_Security_Awareness}.
The questions deal with individual abilities to detect weaknesses in code and skills to deal with these weaknesses. This part of the questionnaire draws from the first author's experience in industrial software engineering and various serious games designed and played to train software developers in secure coding.

The third section in the questionnaire presents questions to measure the intent to comply to secure coding guidelines.
The theoretical background of the questions is given by Bulgurcu et al. \cite{bulgurcu2010information} and Moody et al. \cite{moody2018toward}.
The following six constructs were selected:  {\it self efficacy to comply} (SE-C), {\it intention to comply with the policy} (ITC),  {\it general information security awareness} (GISA), {\it awareness} (ISPA), {\it response cost} (RespCost4) and {\it facilitating conditions} (FacCond5).

The fourth section of the questionnaire is about the factors that influence compliance with secure coding guidelines.
Here we have used two different theories: neutralization theory and security-related stress theory.
For Neutralization Theory (NT), six constructs were selected: defense of necessity (N-DON3), appeal to higher loyalties (N-ATHL1), denial of injury (N-DOI1, N-DOI2, N-DOI3), denial of responsibility (N-DOR3), condemnation of the condemners (N-COC1, N-COC2) and metaphor of the ledger (N-MOTL1).
For Security-Related Stress (SRS), the following three constructs were selected: complexity (CX2, CX4), overload (OL1, OL4), and uncertainty (UC1, UC4)

These constructs were selected based on the industry experience of the first author and practical limitations on the number of questions (to increase response rate).
Additionally, based on lessons learned from teaching secure coding in the industry, we have included additional questions.
These questions are of the following two types: Company Background (CBG) and Background Knowledge (BGK).
The CBG questions intend to assess the established processes in the company related to secure coding guidelines.
The BGK questions intend to assess background knowledge related to secure coding guidelines by the software developer.

\begin{spacing}{.9}
  \begin{table*}[http]
    \renewcommand{\arraystretch}{1.62}
    \scriptsize
    \centering
    \caption{Final Questionnaire Constructs and Adapted Questions}
    \resizebox{\textwidth}{!}{
    \begin{tabular}{|p{0.6cm}|p{0.55cm}|p{1.1cm}|p{1.2cm}|p{12.5cm}|}
      \hline
      \textbf{~~RQ.} &
      \textbf{Type}    &
      \textbf{Ref.} &
      \textbf{Construct} &
      \textbf{~~~~~~~~~~~~~~~~~~~~~~~~~~~~~~~~~~~~~~~~~~~~~~~~~~~~~Survey Question}
      \\ \hline
      \hline
      
      \multirow{9}{*}{~~RQ1}&
      \multirow{5}{*}{CBG} &
      \multirow{7}{*}{~~~~~\textemdash} &
      \textit{CBg1} &
      In my company compliance to secure code guidelines is being checked in projects I work in
      \\ \cline{4-5}
      &
      &
      &
      {\bf \textit{CBg2} } &
      {\bf I know the secure software development lifecycle in my company}
      \\ \cline{4-5}
      &
      &
      &
      \textit{CBg3} &
      To which extent do you work with the $\rule{1cm}{0.15mm}$ secure coding standard? 
      \\ \cline{4-5}
      &
      &
      &
      {\bf \textit{CBg7}} &
      {\bf How is the compliance to secure coding guidelines checked in my current project?}
      \\ \cline{4-5}
      &
      &
      &
      \textit{CBg8} &
      In my company we use a well established secure software development life-cycle
      \\ \cline{2-2} \cline{4-5}
      &
      \multirow{2}{*}{BGK} &
      &
      {\bf \textit{BgK1}} &
      {\bf Compliance to secure coding guidelines is an important part of the development of our products}
      \\ \cline{4-5}
      &
      &
      &
      \textit{BgK2} &
      Which of the following secure coding standards and best practices do you know? 
      \\ \cline{2-5}
      &
      \multirow{2}{*}{PC}&
      ~~~~~\cite{bulgurcu2010information}&
      {\bf ISPA} &
      {\bf I know that my company has a policy that mandates the usage of secure coding guidelines in software development}
      \\ \cline{3-5}
      &
      &
      ~~~~~\cite{moody2018toward}&
      FacCond5 &
      Support is available if I experience difficulties in complying with secure coding guidelines
      \\ \hline
      \hline
      

      \multirow{31}{*}{~~RQ2}&
      \multirow{3}{*}{CBG}&
      \multirow{7}{*}{~~~~~\textemdash} &
      \textit{CBg4} &
      Could you explain why you use secure coding guidelines when writing code for the product you currently develop? 
      \\ \cline{4-5}
      &
      &
      &
      \textit{CBg5} &
      Could you tell us why you do not use secure coding guidelines?
      \\ \cline{4-5}
      &
      &
      &
      \textit{CBg6} &
      Why is compliance to secure coding guidelines not actively being checked in the projects I work in?
      \\ \cline{2-2}\cline{4-5}
      &
      \multirow{11}{*}{PC} &
      &
      \textit{PC-Conf} &
      Complying to SCG makes me feel more confident about the security of the code that I write
      \\ \cline{4-5}
      &
      &
      &
      {\bf \textit{PC-NT}} &
      {\bf In my opinion, to write secure code, I have the necessary time}
      \\ \cline{4-5}
      &
      &
      &
      {\bf \textit{PC-NR}} &
      {\bf In my opinion, to write secure code, I have the necessary resources}
      \\ \cline{4-5}
      &
      &
      &
      {\bf \textit{PC-NF}} &
      {\bf In my opinion, to write secure code, I have the necessary freedom}
      \\ \cline{3-5}
      &
      &
      \multirow{5}{*}{~~~~~\cite{bulgurcu2010information}}  &
      {\bf SE-C1} &
      {\bf In my opinion, to write secure code, I have the necessary skills}
      \\ \cline{4-5}
      &
      &
      &
      {\bf SE-C2} &
      {\bf In my opinion, to write secure code, I have the necessary knowledge}
      \\ \cline{4-5}
      &
      &
      &
      {\bf SE-C3} &
      {\bf In my opinion, to write secure code, I have the necessary  competency}
      \\ \cline{4-5}
      &
      &
      &
      {\bf ITC} &
      {\bf I intend to always comply with secure coding guidelines}
      \\ \cline{4-5}
      &
      &
      &
      {\bf GISA} &
      {\bf I am aware of the existing security threats to the products of my company}
      \\ \cline{3-5}
      &
      &
      ~~~~~\cite{moody2018toward} &
      {\bf RespCost4} &
      {\bf Secure coding guidelines make the task of writing software more difficult}
      \\ \cline{2-5}
      &
      \multirow{11}{*}{NT} &
      \multirow{9}{*}{~~~~~\cite{siponen2010neutralization}} &
      {\bf N-DON3} &
      {\bf It's OK to disregard secure coding guidelines when this means that I deliver my work-packages faster}
      \\ \cline{4-5}
      &
      &
      &
      {\bf N-ATHL1} &
      {\bf It's OK to disregard secure coding guidelines when I would otherwise not get my job done}
      \\ \cline{4-5}
      &
      &
      &
      {\bf N-DOI1} &
      {\bf It's OK to disregard secure coding guidelines when this would result in no harm to the customer}
      \\ \cline{4-5}
      &
      &
      &
      N-DOR3 &
      It's OK to disregard secure coding guidelines if you do not understand them
      \\ \cline{4-5}
      &
      &
      &
      N-DOI2 &
      It's OK to disregard secure coding guidelines if no damage is done to the company
      \\ \cline{4-5}
      &
      &
      &
      N-COC1 &
      It's not as wrong to ignore secure coding guidelines that are not reasonable
      \\ \cline{4-5}
      &
      &
      &
      N-COC2 &
      It's not as wrong to ignore secure coding guidelines that require too much time to comply with
      \\ \cline{4-5}
      &
      &
      &
      N-MOTL1 &
      I feel my general adherence to secure coding guidelines compensates for occasionally them
      \\ \cline{3-5}
      &
      &
      \multirow{3}{*}{~~~~~\textemdash}  &
      {\bf \textit{NT-MArc}} &
      {\bf It’s OK to disregard secure coding practices when this would lead to major architectural changes}
      \\ \cline{4-5}
      &
      &
      &
      \textit{NT-CH} &
      It's OK to disregard secure coding guidelines when this means that it makes my customers happy
      \\ \cline{4-5}
      &
      &
      &
      \textit{NT-SC} &
      It's OK to disregard secure coding guidelines if the software is not safety critical
      \\ \cline{2-5}
      &
      \multirow{6}{*}{SRS}  &
      \multirow{6}{*}{~~~~~\cite{d2014understanding}}&
      CX2 &
      I find that new employees often know more about secure coding than I do
      \\ \cline{4-5}
      &
      &
      &
      CX4 &
      I often find it difficult to understand my organization’s security coding guidelines
      \\ \cline{4-5}
      &
      &
      &
      OL1 &
      Complying to secure coding guidelines force me to do more work than I can handle
      \\ \cline{4-5}
      &
      &
      &
      OL4 &
      I am forced to change my work habits to adapt to my organization’s secure coding guidelines
      \\ \cline{4-5}
      &
      &
      &
      UC1 &
      There are constant changes in secure coding guidelines my organization
      \\ \cline{4-5}
      &
      &
      &
      UC4 &
      There are constant changes in security-related technologies in my organization
      \\ \hline
      \hline


      \multirow{6}{*}{~~RQ3}&
      \multirow{3}{*}{BGK}&
      \multirow{3}{*}{~~~~~\textemdash}&
      \textit{BgK4} &
      What other weaknesses do you pay attention to in developing software for the product you currently work for?
      \\ \cline{4-5}
      &
      &
      &
      \textit{BgK5*} &
      I know about this vulnerability
      \\ \cline{4-5}
      &
      &
      &
      \textit{BgK3} &
      I am aware of negative consequences resulting from exploiting vulnerabilities on the products I deliver software for
      \\ \cline{2-5}
      &
      \multirow{3}{*}{AW}  &
      \multirow{3}{*}{~~~~\cite{2014_Benenson_Defining_Security_Awareness}} &
      \textit{Per1*} &
      I can recognize code that contains this weakness
      \\ \cline{4-5}
      &
      &
      &
      \textit{Be1*} &
      I know how to write code that does not contain this weakness
      \\ \cline{4-5}
      &
      &
      &
      \textit{Prot1*} &
      I understand the possible consequences that can result from exploiting this weakness
      \\ \hline
    \end{tabular}}
    {\\\hspace{\textwidth}
                     {\scriptsize {\bf RQ.}: Research Question, {\bf CBG}: Company Background, {\bf BGK}: Participant Background Knowledge, {\bf PC}: Policy Compliance Theory, {\bf NT}: Neutralization Theory,~~~~~~~~~~~~~~~~~~~~~~\\ {\bf SRS}: Security-Related-Stress Theory, {\bf AW}: Awareness, Note: constructs marked with * are relative to specific software weaknesses~~~~~~~~~~~~~~~~~~~~~~~~~~~~~~~~~~~~~~~~~~~~~~~~~~~~~~~~\\
                     Note: results for the highlighted constructs and questions, which were obtained during the preliminary survey, are presented in the results section
                     ~~~~~~~~~~~~~~~~~~~~~~~~~~~~~~~~~~~
                     }}
    \label{tab:survey:questions}
  \end{table*}
\end{spacing}

\vspace{.3em}

The questions corresponding to RQ1 were included in the first section, and the ones corresponding to RQ3 were included in the second section of the questionnaire.
The questions corresponding to RQ2 were split into the third and fourth sections of the survey, as detailed above.
Company background questions related to RQ2 were included in the third section of the questionnaire.

\subsection{Survey Results}

\begin{table*}[http]
  \renewcommand{\arraystretch}{1.72}
  \scriptsize
  \centering
  \caption{Results of Preliminary Survey Based on a Subset of Survey Questions}
  \begin{tabular}{|p{0.5cm}|p{0.5cm}|p{1.5cm}|p{4cm}|p{0.5cm}|p{0.5cm}|p{0.5cm}|p{0.5cm}|p{0.5cm}|}
   \hline
   {\bf RQ.}                         &
   {\bf Type}                        &
   {\bf ~~Construct}                 &
   {\bf ~~~~~~~~Summary of Question} &
   {\bf ~~SD}  &
   {\bf ~~~D}  &
   {\bf ~~~N}  &
   {\bf ~~~A}  &
   {\bf ~~SA} \\
  \hline
  \hline
  \multirow{4}{*}{~RQ1} &
  \multirow{2}{*}{~CBG} &
  \textit{CBg2} &
  known S-SDLC in the company &
               3.8  &
              15.4  &
              50.0  &
              28.8  &
               1.9 \\
  \cline{3-9}
  &
  &
  \textit{CBg7} &
  SCG being actively checked &
               3.8  &
              21.2  &
              38.5  &
              32.7  &
               3.8 \\
  \cline{2-9}
  &
  ~BGK&
  \textit{BgK1} &
  importance of SCG&
               0.0  &
               3.6  &
              10.9  &
              69.1  &
              16.4  \\
  \cline{2-9}
  &
  ~~~PC &
  ISPA &
  know policies &
               3.6  &
              14.3  &
              33.9  &
              39.3  &
               8.9  \\
  \hline
  \hline
  \multirow{13}{*}{~RQ2}&
  \multirow{9}{*}{~~~PC}&
  \textit{PC-NT} &
  I have necessary time &
               0.0  &
              18.5  &
              24.1 &
              50.0  &
               7.4 \\
  \cline{3-9}
  &
  &
  \textit{PC-NR} &
  I have necessary resources &
               0.0  &
              13.2  &
              32.1 &
              52.8 &
               1.9 \\
  \cline{3-9}
  &
  &
  \textit{PC-NF} &
  I have necessary freedom &
               1.9  &
              11.1  &
              11.1 &
              57.4  &
              18.5 \\
  \cline{3-9}
  &
  &
  SE-C1 &
  I have necessary skills &
               0.0  &
               7.5  &
              32.1 &
              50.9  &
               9.4 \\
  \cline{3-9}
  &
  &
  SE-C2 &
  I have necessary knowledge &
               0.0  &
              13.0  &
              33.3 &
              46.3  &
               7.4 \\
  \cline{3-9}
  &
  &
  SE-C3 &
  I have necessary competencies &
               0.0  &
               7.7  &
              40.4 &
              44.2 &
               7.7 \\
  \cline{3-9}
  &
  &
  ITC &
  intent to comply &
               0.0  &
               1.8  &
              21.4  &
              57.1  &
              19.6  \\
  \cline{3-9}
  &
  &
  GISA &
  know security threats &
               0.0  &
              14.3  &
              25.0  &
              46.4  &
              14.3  \\
  \cline{3-9}
  &
  &
  RespCost4 &
  job more difficult &
               0.0  &
              30.9  &
              30.9  &
              36.4  &
               1.8  \\
  \cline{2-9}
  &
  \multirow{4}{*}{~~~NT} &
  N-DON3 &
  deliver faster &
              24.1  &
              51.9  &
              14.8  &
               9.2  &
               0.0  \\
  \cline{3-9}
  &
  &
  N-ATHL1 &
  not get job done &
              18.5 &
              38.9  &
              33.3  &
               9.3  &
               0.0  \\
  \cline{3-9}
  &
  &
  N-DOI1 &
  no harm to customer &
              14.5  &
              25.5  &
              27.3  &
              21.8  &
              10.9  \\
  \cline{3-9}
  &
  &
  \textit{NT-MArc} &
  major architectural changes &
              18.5  &
              33.3  &
              29.6  &
              13.0  &
               5.6  \\
  \hline
  \end{tabular}
  \renewcommand{\arraystretch}{1.68}
  \begin{tabular}{|p{0.7cm}|p{0.7cm}|p{0.7cm}|}
  \hline
  {\bf ~SD+D}  & 
  {\bf ~~~N}   &
  {\bf ~A+SA}  \\
  \hline
  \hline
    19.2 &
    50.0 &
    30.7 \\
  \hline
    25.0 &
    38.5 &
    36.5 \\
  \hline
     3.6 &
    10.9 &
    85.5 \\
  \hline
    17.9 &
    33.9 &
    48.2 \\
  \hline
  \hline
    18.5 &
    24.1 &
    57.4 \\
  \hline
    13.2 &
    32.1 &
    54.7 \\
  \hline
    13.0 &
    11.1 &
    75.9 \\
  \hline
     7.5 &
    32.1 &
    60.3 \\
  \hline
    13.0 &
    33.3 &
    53.7 \\
  \hline
     7.7 &
    40.4 &
    51.9 \\
  \hline
     1.8 &
    21.4 &
    76.7 \\
  \hline
    14.3 &
    25.0 &
    60.7 \\
  \hline
    30.9 &
    30.9 &
    38.2 \\
  \hline
    76.0 &
    14.8 &
     9.2 \\
  \hline
    57.4 &
    33.3 &
     9.3 \\
  \hline
    40.0 &
    27.3 &
    32.7 \\
  \hline
    51.8 &
    29.6 &
    18.6 \\
  \hline
  \end{tabular}
  {\\\hspace{\textwidth}
                     {\scriptsize {\bf RQ}: Research Question, {\bf CBG}: Company Background, {\bf PC}: Background Knowledge, {\bf PC}: Policy Compliance Theory, ~~~~~~~~~~~~~~~~~~~~~~~~~~~~~~~~~~~~~~~~~~~~~~~~~~~\\{\bf NT}: Neutralization Theory, {\bf SD}: Strongly Disagree, {\bf D}: Disagree, {\bf N}: Neutral, {\bf A}: Agree, {\bf SA}: Strongly Agree}~~~~~~~~~~~~~~~~~~~~~~~~~~~~~~~~~~~~~~~~~~~~~~~~~~~~~~~~~~~~~~~ \\
                     Results show the average percentage of agreement to each individual likert scale point ~~~~~~~~~~~~~~~~~~~~~~~~~~~~~~~~~~~~~~~~~~~~~~~~~~~~~~~~~~~~~~~~~~~~~~~~~~~~~~~~~~~~
                     }
  \label{tab:results:summary}
\end{table*}

Table~\ref{tab:results:summary} shows the results of the preliminary analysis of the survey pilot.
This analysis focuses on the questions that were not substantially changed in the last survey design iteration and made to the final version of the survey.
Since RQ3 was formulated due to the refinements made in the post-review phase, this analysis focuses solely on RQ1 and RQ2.

\subsubsection{Preliminary results for RQ1}
The results in Table~\ref{tab:results:summary} show that, since the majority of the software developers have a neutral opinion and 19.2\% disagree on CBg2 (software developers know about the secure software development lifecycle used in the company they work for), we can conclude that they generally lack knowledge about the company's secure software development lifecycle.
Additionally, there is an indicator that compliance with SCG is being checked (36.5\%); however, the large number of neutral results also indicates that this might be an issue.
Also, the vast majority of software developers recognize the importance of SCG (85.5\%).
Finally, although 48.2\% agree on knowing the secure coding guideline policies (policy compliance, ISPA construct), a large number does not have an opinion (33.9\%) or disagrees on this fact (17.9\%). Therefore we conclude that many software developers lack awareness about secure coding policies since the average agreement is low.
We observe that, although software developers lack awareness of secure software development practices, 85.5\% claim that they are aware of their importance (background knowledge, BgK2 construct).

\subsubsection{Preliminary results for RQ2}
In terms of the factors that lead software developers to use secure coding guidelines (RQ2), the results in Table~\ref{tab:results:summary} can be summarized as follows.
Software developers express having enough freedom (74.9\%), skills (60.3\%), time (57.4\%), resources (54.7\%), knowledge (53.7\%), and competencies (51.9\%) to write secure code according to SCG.
This high agreement values for freedom, time, and resources indicate that executing company processes should not be an issue.
However, when considering the items skills, knowledge, and competencies, these have lower values.
Additionally, software developers express the intention to comply (ITC) with SCG (76.6\%) and also express knowledge about possible threats to the products from the company (60.9\%).
However, software developers are not sure if complying with SCG makes the daily job more difficult.

In terms of SE-C2 (knowledge), the results show that 53.7\% of the software developers do not know secure coding guidelines.
This value correlates very well with the results by Patel et al.~\cite{gitlab_2019}, where they found out that more than 50\% of software developers cannot spot security vulnerabilities in code.

These facts indicate a need for awareness training in secure coding and the application of secure coding guidelines.
Furthermore, the results show that 76.6\% express the intention to comply with secure coding guidelines and their policies and express knowledge about possible threats to the company (60.7\%).
The combined results indicate that secure coding is not a matter of governance but awareness.
Additionally, the results in Table~\ref{tab:results:summary} show that software developers are not sure (30.9\%) if complying with SCG makes the daily job more difficult (38.2\%) or not (30.9\%).
These results can be interpreted as a lack of awareness and, therefore, lack of good sense to evaluate the task's difficulty.

In terms of the factors that lead software developers not to comply with secure coding guidelines, the neutralization theory results are shown in the following.
The majority of software developers (76\%) disagree that SCG should be overlooked to deliver software faster.
Although there is a higher uncertainty than in the previous case, software developers also disagree that secure coding guidelines should not be ignored to get a job done. However, 33\% have an ambiguous opinion on this matter, which is surprising since most software developers agree on the importance of SCG.
Although still with 51.8\% agreement by the developers, the numbers express a higher uncertainty on disregarding SCG if these mean significant architectural changes.
Finally, software developers do not have a definite opinion (27.3\%) if they may ignore secure coding guidelines in case the customer of the software would not be harmed. Furthermore, the agreement level (32.7\%) and disagreement level (40.0\%) show only a slight tendency towards disagreement.
These observations are surprising given that software developers should not ignore secure coding guidelines based on their judgment, e.g., on how the end-customer will use software.

In general, we conclude that software developers agree on not ignoring SCG (i.e., following them). However, considering the agreement level on competencies and skills, software developers might lack the skills to judge whether they comply with the secure coding guidelines.

In a similar study conducted by Assal et al.~\cite{ Assal2019}, called "Think Secure, " 123 software developers from different industry sectors, mostly located in Canada and the United States, were surveyed. The average age of the software developers was 41.3 years, which is comparable with the age range of the participants from our study.
We have looked at the questions and results by Assal et al.  To compare their results to our own, the results presented in table \ref{tab:results:summary} were further processed. First, we use the following standard Likert mapping: \textit{strongly disagree} $\rightarrow$ 1, ... \textit{strongly agree} $\rightarrow$ 5. Also, two of the Likert scales are inverted: the construct PC-NT (\textit{have necessary time}) is inverted to match the question in  Assal et al. and similarly for the construct SE-C2 (\textit{have knowledge}) as compared to D24 (\textit{have no knowledge}). We also considered a mapping between a compound construct based on the average agreement between CBg2 and CBg7 to the construct \textit{we have security procedures} in the survey by Assal et al.

\begin{spacing}{.9}
  \begin{table*}[htb]
    \renewcommand{\arraystretch}{1.37}
    \footnotesize
    \centering
    \caption[caption]{Comparison between this study and the study from Assal et al.}
    \begin{tabular}{|p{2cm}|p{2cm}|p{4cm}|p{2cm}|}
      \hline
      \multicolumn{2}{|c|}{This study}   &
      \multicolumn{2}{|c|}{"Think Secure" Study \cite{Assal2019}} 
      \\ \hline
      \multicolumn{1}{|c|}{Construct} &
      \multicolumn{1}{|c|}{Average}   &
      \multicolumn{1}{|c|}{Construct} &
      \multicolumn{1}{|c|}{Average}
      \\ \hline
      \hline
      CBg2 + CBg7 $^{\dagger}$ & \multicolumn{1}{|c|}{3.1 $\pm$ 0.6}   & we have security procedures & \multicolumn{1}{|c|}{3.8 $\pm$ 0.9}
      \\ \hline
      BkG1        & \multicolumn{1}{|c|}{4,0 $\pm$ 0,3}   & security is important       & \multicolumn{1}{|c|}{4.3 $\pm$ 0.2}
      \\ \hline
      PC-NT$^*$      & \multicolumn{1}{|c|}{2.5 $\pm$ 0.8}   & D23: no time                & \multicolumn{1}{|c|}{2.3 $\pm$ 0.3}
      \\ \hline
      SE-C2$^*$      & \multicolumn{1}{|c|}{2.5 $\pm$ 0.7}   & D24: no knowledge           & \multicolumn{1}{|c|}{2.6 $\pm$ 0.3}
      \\ \hline
    \end{tabular}
    \label{tab:survey:compare}
    \\\hspace{\textwidth}
    {\scriptsize
      Note$^{\dagger}$: Construct corresponds to the average of CBg2 and CBg7 from table~\ref{tab:results:summary} ~~~~~~~~~~~~~~~~~~~~~~~\\
      Note$^*$: Construct has an inverted Likert scale in comparison to table~\ref{tab:results:summary} ~~~~~~~~~~~~~~~~~~~~~~~~~~~~~
    }
  \end{table*}
\end{spacing}

\vspace{.5em}
Table~\ref{tab:survey:compare} shows a summary of the comparison of the average agreement between the current work and the survey from Assal el al. Note that the errors in the "think survey," except for the construct (CBg2+CBg7), are smaller than in our survey - 
this fact is not a surprise, given that the "think survey" includes 140 participants compared with 59 in our pilot study survey.
Table~\ref{tab:survey:compare} shows that results concerning the existence of security procedures, the knowledge of the importance of security, and software developers' facilitating conditions are comparable between the two surveys.
Furthermore, the alignment of the results validates our questionnaire, the results of the data analysis and approach.

\subsection{Threats to validity}
A threat to our results' validity steams from the fact that, although a reasonable number of answers were collected (59), data comes from a pilot study. To improve the results' quality, we did not consider the full collection of questions and answers in this study from the pilot survey - only those that also survived the post-review process unchanged. Also, the data collected in this survey comes mainly from companies based in Asia; regional results might vary and may lead to different conclusions. However, we notice that the results that we have obtained in our survey agree with the work from Assal et al. \cite{Assal2019}, which was performed in the United States and Canada and was also deployed over several different companies.

\subsection{Impact of this work}
More effort and different measures are needed to increase code security, particularly for software used in critical infrastructures. Our preliminary findings indicate that software developers have the resources for writing secure code, but not necessarily the knowledge, capabilities, and skills. It needs to be discussed to what extent software developers can judge whether they comply with secure coding guidelines. More research and a more in-depth analysis of data are necessary to resolve the dichotomy between the current status-quo and the number of vulnerabilities vs. software developer self-assessment on the secure coding guideline topic. 
A better understanding of the topic may contribute to more secure code and more effective and efficient organizational structures for secure coding. The first data analysis of preliminary data and the publication of our questionnaire contribute herein.
\section{Conclusions}
\label{sec:conclusions}
The work presented in this paper is motivated by industry needs.
Secure coding is a fundamental competence that every software developer working in the industry should have.
Being competent in secure coding and secure coding guidelines can significantly impact the industry in terms of product quality and security.
This impact becomes especially significant for software that is deployed in critical infrastructures.
However, software vulnerabilities are still increasing, which raises the question of - why is it so?

The survey developed in this work tries to answer this question by focusing on the human factor - the software developer - and its compliance to secure coding guidelines. In particular, we explore the following three aspects: which factors lead software developers to use or ignore secure coding guidelines, how well established are secure programming practices in the industry, and how well do software developers know secure coding guidelines.
Our survey design is based on a mixture of previously well-established theories, the authors' industry experience, and lessons learned from previous serious games on secure software development and secure coding guidelines.

The survey was piloted in three different companies during several CyberSecurity Challenges workshops, which are secure coding workshops to train software developers in the industry on secure software development.
The preliminary results, which are partially confirmed by a similar previous work, show that, while software developers express intention to comply with secure coding guidelines, real-world knowledge on the guidelines is lacking.
Our results also indicate the need for running secure coding awareness campaigns directed towards software developers in the industry.
According to the authors' many years of experience in the industry, software developers tend to overestimate their secure coding capabilities.
Together with the results hereby presented, we hypothesize that software developers need to be challenged to grow in terms of knowledge and awareness on secure coding.

We believe that CTF-like serious games on secure coding, which are designed for software developers in the industry, are an adequate method to raise secure coding awareness.
Awareness training based on these kinds of games poses secure coding challenges to players, which can be used to measure the self-knowledge on secure coding and raise awareness on secure coding guidelines.

Based on a large scale deployment of the designed survey, further work will extract practical advice for CTF-like awareness campaigns and analyze the relationship between the different variables to gain further insight.
One such additional insight might be, e.g., the relationship between the number of years of experience of participants and their knowledge level of secure coding guidelines or willingness to use secure coding guidelines.

\section*{Supporting Data}
The raw data collected in the CyberSecurity Challenge workshops, which is the basis for this work, is openly available in the Zenodo~\cite{tiago_espinha_gasiba_2020_4075282} platform.
The raw survey data is provided in Comma Separated Values (CSV) format.

\section*{Acknowledgements}
The authors would like to thank the survey participants for completing the survey, their time, and their comments. The authors would also like to thank the three anonymized companies which have allowed the survey pilot to be carried and the manuscript reviewers.
The authors would also like to thank Kristian Beckers and Thomas Diefenbach for their helpful, insightful, and constructive comments and discussions.

This work is co-financed by portuguese national funds through FCT - Fundação para a Ciência e Tecnologia, I.P., under the project FCT UIDB/04466/2020. Furthermore, the third author thanks the Instituto Universitário de Lisboa and ISTAR-IUL, for their support.

\bibliographystyle{IEEEtran}
\bibliography{bibliography}

\end{document}